\documentclass{ws-mpla}
\usepackage[super]{cite}
\usepackage{graphicx}
\usepackage[utf8x]{inputenc}
\usepackage[english]{babel}
\usepackage{graphicx}
\usepackage{float}
\usepackage{multirow}
\usepackage{cite}
\usepackage{setspace, longtable, amstext}
\usepackage{selinput}
\usepackage{hyperref}

\usepackage{amssymb}
\usepackage{amsmath}

\usepackage{amsopn}
\def\beq{\begin{equation}}
\def\eeq{\end{equation}}
\def\bea{\begin{eqnarray}}
\def\eea{\end{eqnarray}}

\begin{document}

\markboth{Authors' Names}
{Instructions for Typing Manuscripts (Paper's Title)}

\catchline{}{}{}{}{}

\title{Dark halos and Tully-Fisher relation testing modified gravity}

\author{A. Amekhyan} 
\author{S. Sargsyan} 
\address{Center for Cosmology and Astrophysics, Alikhanian National Laboratory and Yerevan State University, Yerevan, Armenia
}

\maketitle

\begin{history}
\received{... 2023}
\accepted{...}
\end{history}

\begin{abstract}
We involve the galactic halo observational data to test the weak field General Relativity involving the cosmological constant.  Using the data for 15 hydrogen (Hi) VLA super spirals and the Tully-Fisher relation we obtain constraints for each galaxy. The results are consistent with previous results for spiral galaxies, as well as with the scaling relations for the halos, thus confirming the efficiency of the use of dark halo data and Tully-Fisher relation, including the baryonic Tully–Fisher index (BTFR), for testing modified gravity models.
\end{abstract}

\keywords{Galactic halos.}

\ccode{PACS numbers: }

\section{Introduction}

General Relativity(GR) is the fundamental theory enabling to describe the key properties of the evolving Universe. At the same time the revealing of the dark sector of the Universe, as well as the recent appearance of the cosmological tensions, such as the Hubble tension \cite{R1,Val,R2}, essentially stipulates the studies on the extensions of General Relativity and modified gravity theories, e.g. \cite{Kop,Dai1,Dai2,Bo,Cap}. Among such studies is the weak-field modification of GR which naturally includes the cosmological constant $\Lambda$ as a fundamental constant \cite{G1,G2,GS2,GS3}. The metric tensor components then have the form $(c=1)$ \cite{GS2,GS3}
\begin {equation}
g_{00} = 1 - \frac{2 G m}{r} - \frac{\Lambda r^2}{3}; 
\qquad g_{rr} = \left(1 - \frac{2 G m}{r} - \frac {\Lambda r^2}{3}\right)^{-1}.
\end {equation}

Also, it is important to note that, this approach is shown to be able to describe the dynamics of galaxy groups and clusters, the Hubble tension as a result of two galactic flows, local and global ones, with not coinciding Hubble parameters \cite{GS4,GS5}. 
It was also shown that dark halos, including those traced by Cosmic Microwave Background data (e.g. \cite{DEP1,DEP2,AS1}), can be used for obtaining constraints on the modified gravity models, namely, regarding such the halo parameters as the sparsity\cite{AS2} and scaling relations\cite{AS3}.
 
Existence of dark matter (DM) proposed by Zwicky \cite{Zwicky} based on the estimations of the mass of the Coma cluster and the kinematics of its galaxies, has found strong support later by Rubin \cite{Rubin} while studying the rotation curves of the spiral galaxies. Currently, the flatness of the galactic rotation curves is being considered as a key indication of the huge amount of DM at large galactic radii, but also is a subject of studies within modified gravity models. There are numerous candidates of DM particles, both baryonic \cite{Bai} (MACHOs, primordial black holes) and non-baryonic \cite{Ghosh} (neutrinos, WIMPs, axions, etc) having still no indications in the attempts of their detection in dedicated experiments. Among the approaches to probe the nature of DM is the study of the properties of the galactic dark halos such as the halo splashback radius \cite{ONeil}, indicating halo's boundary, as well as kinematic phenomena around halos of their dynamical evolution \cite{Luo}.  MOND (see e.g. \cite{Bekenstein}), F(R) gravity \cite{Zaregonbadi} and other models are used to describe the flat rotation curves of galaxies, galactic flows, the cosmological tensions \cite{Bo,Cap}.  Therefore, the use of possibly diverse observational data to trace the modified gravity models is of particular importance. In this paper we continue the previous studies of the dark halos \cite{AS2,AS3} based on the modified weak field GR model and aim to obtain constraints on the value of the cosmological constant using the halo mass function and Tully-Fisher relation.

\section{DM halo mass function}

The halo mass function is one of the fundamental concepts in large scale structure formation theory (e.g. \cite{Gupta} and references therein). It  describes the number density of halos in given volume
\begin {equation} \label{mod}
\frac{dn}{dM} =  \frac{\rho_m}{M^2}{F(\sigma)}\Big|\frac{dln\sigma}{dlnM}\Big|,
\end {equation}
where
\begin{equation}
\sigma^{2}(R_L,z)=\frac{1}{2\pi^2} \int_{0}^{\infty} k^{2}W^{2}(kR_L)P(k,z) \,dk\
\end{equation}
is the variance of the linear density field. $P(k,z)$ is the matter power spectrum and 
\begin{equation}
R_L=\Bigl(\frac{3M}{4\pi\rho_m}\Bigl)^\frac{1}{3}.
\end{equation}
$F(\sigma)$ is another important quantity, which describes the mass fraction in the collapsed volume
\begin{equation}
F(\sigma)=\sqrt{\frac{2}{\pi}}\frac{\delta_c}{{\sigma}}\exp\Bigl(-\frac{\delta_c^2}{2\sigma^2}\Bigl).
\end{equation}
In the $\Lambda$-gravity context, instead of $R_L$, we can define a critical distance scale for a system, where the repulsive $\Lambda$-term becomes dominant over the Newtonian gravity \cite{GS3,GS4}
\begin{equation}
r_{crit}^3=\frac{3GM}{\Lambda}.
\end{equation}
Then combining these two equations, one has for $\Lambda$
\begin{equation}
\Lambda=\frac{3G\rho_m}{\frac{dn}{dlogM}}\frac{F(\sigma)}{r_{crit}^3}\frac{dlog(\sigma)}{dlogM}.
\end{equation}
The quantities $\frac{dn}{dlog(M)}$, $\frac{dlog(\sigma)}{dlog(M)}$ and multiplicity function are fixed via N-body simulations \cite{Gupta}.

As a result, for $\Lambda$ we obtain the value $0.33\times10^{-52}m^{-2}$. Note, that	the current value of the cosmological constant provided by Cosmic Microwave Background (CMB) data is $\Lambda\simeq 1.1\times10^{-52}m^{-2}$ \cite{CMB}. Thus, through galactic halo mass function we obtain the cosmological constant's value, which is consistent with previous results \cite{AS2,AS3}.
	
\section{Tully-Fisher relation and $\Lambda$-gravity}

The Tully-Fisher (TF) relation is the relationship between galaxy mass (luminosity) and rotation velocity \cite{TF} $M \approx V^b$. It is a useful tool for determining galaxy distances, as well as for certain properties of DM. Recently we have analyzed TF relation for super spirals in context of $\Lambda$-gravity \cite{AS2}. The observational data used in \cite{AS2} were possessing TF relation index break at stellar mass $\frac{M_{stars}}{M_{\odot}}>11.5$, while in recent \cite{Teodoro} it is found to be unbroken up to $\frac{M_{stars}}{M_{\odot}}\simeq 11.7$. This is a remarkable observational fact and in this work we use for testing the modified gravity model with a cosmological constant, the data on 15 massive ($M>10^{11}M_{\odot}$) spiral galaxies, with $V_{rot}>300kms^{-1}$ rotational velocity and $d<250Mpc$ distance (Table 1).
Then, the virial theorem with $\Lambda$ term leads to
\begin{equation}
V_c^2=\frac{GM}{r}-\frac{\Lambda c^{2}r^{2}}{3}.
\end{equation}

Using the virial theorem with $\Lambda$ term in TF relation one can obtain the following expression
\begin{equation}
(\frac{V_c-E(V_c)}{V_c})^b \simeq 1-\frac{\Lambda c^{2}r^{3}}{3GM_{baryonic}},
\end{equation}
where the baryonic Tully–Fisher relation (BTFR) index $3.58$ reported in \cite{Teodoro} is used, and $E(V_c)$ is velocity error limit for each galaxy.
Then, using BTFR index we obtain limits for cosmological constant. The results for each galaxy are presented in the Table 1.

\begin{table}\label{tab1}
\centering
\tbl{}
{
\begin{tabular}{ |p{3.2cm}|p{2.4cm}|p{2.4cm}||p{2.4cm}| |p{2.4cm}|}
\hline
\multicolumn{5}{|c|}{ \textbf{Galaxy data vs $\Lambda$}} \\
\hline
Galaxy& $HI$ data & ${M_{stars}}/{M_{\odot}}$  & Distance (Mpc)&$\Lambda$ ($m^{-2}$) $10^{-51}$\\
\hline
NGC0338 &   WHISP & 11.2 & $78 \pm 11$ & 4.57\\
\hline 
NGC1167 &   WHISP & 11.3 & $73 \pm 7$ & 4.45\\ 
\hline
NGC1324 &   SG06 & 11.2 & $72 \pm 10$ & 4.42\\
\hline
NGC2599  &  WHISP & 11.1 & $73 \pm 8$ & 4.41\\
\hline
NGC2713 &    new & 11.4 &$ 68 \pm 9$ & 3.74\\
\hline
NGC2862 &   SG06 & 11.1 & $68 \pm 12$ & 3.77\\
\hline
NGC5440 &    new & 11.1 & $57 \pm 4$ & 3.45\\
\hline
NGC5533 &   WHISP  & 11.0 & $45 \pm 6$ & 3.33\\ 
\hline
NGC5635  &  new  & 11.1 & $72 \pm 13$ & 3.77\\
\hline
NGC5790 &   new  & 11.3 & $164 \pm 15$ & 3.69\\
\hline
UGC02849 &  SG06  & 11.2 & $138 \pm 25$ & 4.38\\
\hline
UGC02885 &  H13  & 11.2 & $71 \pm 9$ & 5.50\\
\hline
UGC08179 &  new & 11.6 & $232 \pm 18$ & 5.15\\
\hline
UGC12591 &  new & 11.6 & $97 \pm 7$ & 5.27\\
\hline
UGC12811 &  new & 11.4 & $174 \pm 13$ & 5.80\\
\hline
\end{tabular} 
}
\end{table}

\section{Conclusions}

The role of the cosmological constant is considered of particular importance in modern cosmology. The weak-field modified GR involving the cosmological constant $\Lambda$ has been previously used to describe the dynamics of galaxy groups and clusters \cite{GS3,GS4}. Therefore the study of its predictions using the observational data on the galactic dark halos can enable to trace that version of modified gravity probing the value of the cosmological constant. In our analysis we used the Tully-Fisher relation for the data of a sample of spiral galaxies. Particularly, the recently evaluated baryonic Tully–Fisher relation (BTFR) index \cite{Teodoro} is used. It is remarkable that the values of $\Lambda$ which we obtain are consistent with the value of the cosmological constant following from CMB data. Thus, our analysis indicates the efficiency of the use of the Tully-Fisher relation for the study for the modified gravity models and the importance of continuation of such studies with samples of higher redshifts.

\section{Acknowledgment}

We are thankful to the referee for helpful comments. This work was made possible in part by a research grant from the Yervant Terzian Armenian National Science and Education fund, ANSEF-PS-astroth-2922.


\newpage
\begin{thebibliography}{17}

\bibitem{R1} A.G. Riess, {\it Nature Review Physics}, 2, 10 (2020)
\bibitem{Val} E. Di Valentino E., et al, {\it Class. Quant. Grav.}, 38, 153001 (2021)
\bibitem{R2} A.G. Riess et al, ApJ, 938, 36 (2022)
\bibitem{Dai1} M. Dainotti et al,  ApJ, 912, 150 (2021) 
\bibitem{Dai2} M. Dainotti  et al,	arXiv:2301.10572 (2023)
\bibitem{Kop} S. Kopeikin, Ed. {\it Frontiers in relativistic celestial mechanics}, Vol. 2 Applications and Experiments. (de Gruyter, 2014)
\bibitem{Bo} F. Bouchè, S. Capozziello, V. Salzano, K. Umetsu,  {\it Eur. Phys. J. C}, 82, 652 (2022)
\bibitem{Cap} S. Capozziello, G. Lambiase,  {\it Eur. Phys. J. Plus}, 137, 735 (2022) 
\bibitem{G1}  V.G. Gurzadyan V.G.,  {\it Observatory}, {\bf 105}, 42 (1985)
\bibitem{G2}  V.G. Gurzadyan, \textit{Eur. Phys. J. Plus} {\bf 134}, 14 (2019)
\bibitem{GS2} V.G. Gurzadyan, A. Stepanian, \textit{Eur. Phys. J. C}, {\bf 78}, 632, (2018)
\bibitem{GS3} V.G. Gurzadyan, A. Stepanian,  \textit{Eur. Phys. J. C}, {\bf 79}, 169 (2019)
\bibitem{Kr} A.V.	Kravtsov, ApJ Lett, 764, L31 (2013)
\bibitem{GS4} V.G. Gurzadyan, A. Stepanian,  {\it Eur. Phys. J. Plus}, 136, 235 (2021)
\bibitem{GS5} V.G. Gurzadyan, A. Stepanian,  {\it A\&A}, 653, A145 (2021)
\bibitem{DEP1} F. De Paolis, et al, {\it A\&A}, 565. L3 (2014)
\bibitem{DEP2} F. De Paolis, et al, {\it A\&A}, 629, A87 (2019) 
\bibitem{AS1} A. Amekhyan, {\it IJMPD}, 28, 2040016 (2019)
\bibitem{AS2} A. Amekhyan, et al, \textit{Modern Physics Letters A}, {\bf 35},  2050295 (2020)
\bibitem{AS3}  A. Amekhyan, et al, \textit{Research in Astronomy and Astrophysics} {\bf 21.12}, 309 (2021)
\bibitem{Zwicky} F. Zwicky, \textit{Helvetica Physica Acta}, {\bf 6}, 110, (1933)
\bibitem{Rubin} V.C. Rubin \textit{et al.}, \textit{ApJ}, \textbf{159}, 379 (1970)
\bibitem{Bai} B. Yang, A. J. Long, S. Lu, \textit{JCAP}, \textbf{2020.09}, 044 (2020)
\bibitem{Ghosh} D.K. Ghosh \textit{et al.}, arXiv:2305.09188, (2023)
\bibitem{ONeil} S. O'Neil, \textit{et al.}, \textit{MNRAS}, \textbf{504}, 4649 (2021)
\bibitem{Luo} X. Luo \textit{et al.}, arXiv:2209.06488 (2022)
\bibitem{Bekenstein} J.D. Bekenstein, \textit{Contemporary Physics}, {\bf 47}, 387 (2006)
\bibitem{Zaregonbadi} R. Zaregonbadi \textit{et al.}, \textit{Phys Rev D}, \textbf{94.8}, 084052, (2016)
\bibitem{Gupta} S. Gupta \textit{et al.}, \textit{Phys Rev D}, \textbf{105}, 043538 (2022)
\bibitem{CMB} Planck Collaboration, VI., {\it A\&A}, 641, A6 (2020)
\bibitem{TF} R.B. Tully, J.R. Fisher, \textit{A\&A}, {\bf 54}, 661 (1977)
\bibitem{Teodoro} E.M. Di Teodoro \textit{et al.}, \textit{MNRAS}, \textbf{518}, 6340 (2023)

\end{thebibliography}
\end{document}